\documentstyle[12pt]{article}
\makeatletter

\ifcase \@ptsize
\or 
\or 
\fi
\advance\textheight by \topskip

\ifcase \@ptsize
\or 
\or 
\fi

\def\section{\@startsection {section}{1}{\z@}{-3.5ex plus -1ex minus
 -.2ex}{2.3ex plus .2ex}{\large\bf\centering}}
\def\subsection{\@startsection{subsection}{2}{\z@}{-3.25ex plus -1ex minus
 -.2ex}{1.5ex plus .2ex}{\sc}}
\def\@cite#1#2{\nolinebreak$^{[\scriptstyle #1\if@tempswa , #2\fi]}$}
\def\@citex[#1]#2{\if@filesw\immediate\write\@auxout{\string\citation{#2}}\fi
  \def\@citea{}\@cite{\@for\@citeb:=#2\do
    {\@citea\def\@citea{,\penalty\@m}\@ifundefined
       {b@\@citeb}{{\bf ?}\@warning
       {Citation `\@citeb' on page \thepage \space undefined}}%
{\csname b@\@citeb\endcsname}}}{#1}}
\@definecounter{footnote}

\gdef\@publabel{\hfil}
\gdef\@pubdate{\null}
\gdef\@pubnumber{\null}
\gdef\@author{\null}
\gdef\@title{\null}
\gdef\@abstract{\null}
\long\def\pubdate#1{\gdef\@pubdate{#1}}
\long\def\pubnumber#1{\gdef\@pubnumber{#1}}
\long\def\publabel#1{\gdef\@publabel{#1}}
\long\def\author#1{\gdef\@author{#1}}
\long\def\title#1{\gdef\@title{#1}}
\long\def\abstract#1{\gdef\@abstract{#1}}
\def\titlerelax{
\let\maketitle\relax
\let\settitleparameters\relax
\let\consolidatetitle\relax
\let\inittitlepage\relax
\let\finishtitlepage\relax
\let\titlepagecontents\relax
\let\multithanks\relax
\let\titlebaselines\relax
\let\@makepub\relax
\let\@maketitle\relax
\let\@makeauthor\relax
\let\@makeabstract\relax
\let\@maketitlenote\relax
\let\thanks\relax
\let\titlerelax\relax}
\def\titleclean
{\gdef\@titlenote{}
\gdef\@abstract{}
\gdef\@author{}
\gdef\@title{}
\gdef\@pubdate{}\gdef\@pubnumber{}\gdef\@publabel{}
\gdef\@dpublabel{}
}

\def\@makepub{\vbox to \z@{\hbox to \textwidth{\hfill
\@publabel \hfill
\llap{\parbox[t]{0.25\textwidth}{\raggedleft\@pubnumber}}}%
\vss}}
\def\@maketitle{\vskip 80pt \begin{center}
 {\LARGE \@title \par}
 \end{center}}
\def\@makeauthor{{\def\and{\smallskip {\normalsize \rm and\smallskip}}
\long\def\address##1{{\def\and{\\and\\}\medskip
				{\small \it \\##1\\}
}}
{\centering
 \vskip 2.5em
 \large \lineskip .75em
 \@author}
 \par}}
\def\@makedate{\vskip 1.5em
 {\raggedright \small \noindent\@pubdate \par}}
\def\@makeabstract{\vskip 1.5em
{\small
\begin{center}
{\bf ABSTRACT\vspace{-.5em}\vspace{0pt}}
\end{center}
\quotation \@abstract \endquotation}}
\def\maketitle{
\let\footnotesize\small \setcounter{page}{1}
\@makepub
\@maketitle
\@makeauthor
\@makeabstract
\@thanks
\@makedate
\setcounter{footnote}{0}
}

\begin{document}
\newcommand{\hsp}{\hspace{0.08in}}
\newcommand{\eqn}{\begin{equation}}
\newcommand{\e}{\end{equation}}
\bibliographystyle{npb}

\pubnumber{hep-th/9505124 \ DAMTP-95-27 \\\today}

\title{Fusion of $SO(N)$ reflection matrices}
\author{N. J. MacKay\footnote{n.j.mackay@damtp.cam.ac.uk}
\address{Dept of Applied Maths and Theoretical Physics, \linebreak
         Cambridge University, \linebreak
         Cambridge, CB3 9EW, \linebreak
         England}}
\abstract{
We examine the reflection matrix acting on the $SO(N)$ vector multiplet
and fuse it to obtain that acting on the rank two particle
multiplet, and give its decomposition.
}

\maketitle
\baselineskip 18pt
\parskip 18pt
\parindent 10pt

\section{Introduction}

There has been much interest lately in the integrability properties
of massive 1+1-D theories on a half-line. However, relatively
little of this has been directed at the construction of explicit
solutions of the reflection equation, the analogue for reflection
($K$-)matrices of the Yang-Baxter equation (YBE) for scattering
($S$-)matrices. In this paper we apply the fusion procedure\cite{MN} to
the reflection matrix in the vector representation of $SO(N)$
to find that in the second particle multiplet, an $SO(N)$-reducible
representation whose components are the second rank antisymmetric
tensor and the singlet. Such $K$-matrices would be expected to
apply to the $SO(N)$ principal chiral model on the half-line
and thus to differ in pole structure from the $SO(N)$ $\sigma$-model
investigated by Ghoshal\cite{Gho}, which is not expected to have bulk
bound states\cite{ZZ}. In carrying this out we start from
the $K$-matrices provided by Cherednik\cite{Ch} and investigate
their properties, comparing the matrix structure with that of
Ghoshal, before fusing them. A proper interpretation of our
results in terms of the integrability (or otherwise) of the
physical boundary condition is, however, lacking.

\section{Reflection matrix in the vector representation}

First let us recall that the $S$-matrix solution of the YBE
for two particles in vector representations of $SO(N)$ is\cite{ZZ}
\eqn\label{S}
S_{11}(\theta) = X_{11}(\theta) \sigma(\theta)\: {\bf P}\,
(P_S + [2]P_A + [2][h]P_0) \;,
\e
where
$$
[a]\equiv {{\theta + {ai\pi\over h}} \over
{\theta - {ai\pi\over h}}} \;,
$$
$h=N-2$ is the dual Coxeter number of $SO(N)$, ${\bf P}$ indicates
transposition of states in a tensor product,
and $P_S$, $P_A$, and $P_0$ are projectors onto the second rank
symmetric traceless and antisymmetric tensors and the singlet
respectively. $\sigma(\theta)$ is a scalar factor chosen
such that with $X_{11}=1$ there are no poles in the physical strip
$0\leq {\rm Im}\,\theta \leq \pi$,

$$
\sigma(\theta)={{\Gamma({1\over2}+\theta/2i\pi)\Gamma({1\over
h}+\theta/2i\pi)\Gamma({1\over 2} + {1\over h} -\theta/2i\pi)
\Gamma(-\theta/2i\pi)}
\over {
\Gamma({1\over2}-\theta/2i\pi)\Gamma({1\over
h}-\theta/2i\pi)\Gamma({1\over 2} + {1\over h} + \theta/2i\pi)
\Gamma(+\theta/2i\pi)}} \;,
$$
and $X_{11}(\theta)$ is a
so-called CDD factor which is used here to give the physical
pole structure. The $SO(N)$ $\sigma$-model is expected to have
no bound states and so has $X_{11}=1$. Here we wish to assume
the existence of a second, fused particle state proportional
to $P_A+P_0$ (so that the second particle multiplet is the
$adjoint\oplus singlet$ representation of $SO(N)$, which is an
irreducible representation of the underlying Yangian charge algebra),
and so take
\eqn\label{X}
X_{11}(\theta) = -(2)(N-4)\;, \hspace{0.5in}{\rm where}\;\;\;
(x)\equiv {{\sinh\left({\theta\over 2} + {xi\pi\over h}\right)}
\over      {\sinh\left({\theta\over 2} - {xi\pi\over h}\right)}} \;.
\e

We next recall some ideas from Cherednik's paper\cite{Ch}.
First, this $S$-matrix, $S\equiv {}_+S$ (dropping for the
moment the `$11$' suffices), in principle only applies when
both particles are right-moving. If the reflection matrix is
$K_1(\theta)$ (where, as with $S$, for the moment we drop the suffix
`$1$') then we also need to define scattering matrices for
one left- and one right-moving particle, ${}_0 S$, and for
two left-moving particles, ${}_-S$. We then have
\begin{eqnarray*}
{}_+S(\theta) & = &
1\otimes K(0) \,.\: {}_0S(\theta) \,.\, 1\otimes K(0) \;\\
{}_-S(\theta) & = &
K(0)\otimes 1 \,.\: {}_0S(\theta) \,.\, K(0)\otimes 1 \;,
\end{eqnarray*}
in terms of which the reflection equation is
\eqn\label{reflect}
{}_-S(\phi-\theta) \,.\, 1\otimes K(\phi) \,.\,
{}_0S(\phi+\theta) \,.\, 1\otimes K(\theta) =
1\otimes K(\theta) \,.\: {}_0S(\phi+\theta) \,.\,
1\otimes K(\phi) \,.\: {}_-S(\phi-\theta) \;.
\e
This $K$ must satisfy both unitarity
\eqn\label{u}
K(\theta)\:K(-\theta) = I
\e
and a combined crossing and unitarity
relation\cite{GhZ}: if we write the $S$-matrix acting on two vectors
with indices $k,l$ as $S_{kl}^{ij}$ and the $K$-matrix acting on a
vector with index $k$ as $K_k^i$, then this is
\eqn\label{cu}
K_i^j(i\pi/ 2-\theta) =\, {}_0S_{kl}^{ij}(2\theta)
\,K_l^k({i\pi/ 2}+\theta) \;.
\e
Note that, in contrast to the diagonal case\cite{toda}, this
relation cannot be rewritten to give the $S$-matrix as a product
of reflection matrices, so that the interpretation of
scattering as being equivalent to the placing of a two-sided
mirror at the point of scattering fails.

Cherednik gave a solution to (\ref{reflect}):
$$
K(\theta) \propto E + c\theta\,I \;,
$$
where $I$ is the $N\times N$ identity matrix, $c$ is an (undetermined)
constant and $E$ some matrix such that $E^2=I$. For $SU(N)$ (which
we can recover by setting the trace operator to zero in the above
$S$-matrix) this is a solution for all $c,E$, whilst in the $SO(N)$
case it is only a solution if
\eqn\label{trace}
c=-{2h\over i\pi\,{\rm\em Trace}E }\;.
\e
(At this stage we should also point out how this and all
subsequent calculations are most easily performed. We use
Brauer's diagrammatic representation\cite{Brauer} for the identity,
transposition and trace operators on two vectors in which
multiplication in the algebra corresponds to concatenation of the
symbols, and which has been used to calculate the fused
$S$-matrices\cite{facsm} acting in $P_A+P_0$. It is then simple
to augment the algebra with $E$, the corresponding symbol
being a bead on one of the strands, with a bead on a loop equalling
${\rm\em Trace}E$ and two beads on the same strand disappearing.)

The given matrix structure of $K$ satisfies (\ref{u}), and
satisfies (\ref{cu}) provided (\ref{trace}) holds. In the $su(N)$
case (\ref{cu}) is more subtle since representations are not
self-conjugate, and we need to use $S_{1\bar{1}}$. Once again,
however, we find that (\ref{trace}) must hold for (\ref{cu}) to be
satisfied, this time with $h=N$. This situation for
$K$-matrices is analogous to that for the crossing parameter $h$
for the $S$-matrices: in the $SO(N)$ case this parameter
was explicit in the $S$-matrix (\ref{S}) (because the vector
representation is self-conjugate) but in the $SU(N)$ case the YBE
was not enough to fix it, and it only appeared after the separate
implementation of crossing symmetry.

Cherednik required the distinction between ${}_+S$, ${}_-S$ and
${}_0S$ because his $K(0)\neq I$. We shall choose instead to work
with $K(\theta)\propto I + c\theta\, E$, so that we can require
the more physical condition $K(0)=I$, with ${}_+S={}_-S={}_0S$.
All of our results apply also to Cherednik's $K$ under the
change $[\;] \mapsto -[\;]$. We
write the solution, with an overall factor to be determined, as
\eqn\label{K}
K(\theta) = \tau(\theta)\; (P^- - [h/ci\pi]P^+) \;,
\e
where
$$
P^\pm \equiv {1\over 2} ( I \pm E )
$$
are projectors. This $K$ solves (\ref{reflect}), and satisfies
(\ref{u},\ref{cu}) provided
\begin{eqnarray}
\nonumber
\tau(\theta)\:\tau(-\theta) & = & 1 \\[0.2in] \label{tau}
{{\tau({i\pi\over 2} - \theta)} \over {\tau({i\pi\over 2} + \theta)}}
& = &  X_{11}(2\theta)\sigma(2\theta)
\,\left[{h\over 2}\right]\,\left[{h\over ci\pi}-{h\over 2}\right]\;,
\\ \nonumber
\end{eqnarray}
which is solved by
\begin{eqnarray*}
\tau(\theta) & = & -(1-N/2)(-N/2)(3-N)\;
{{
\Gamma({1\over4}+\theta/2i\pi)\Gamma({1\over 4}+{1\over
2h}-\theta/2i\pi)\Gamma({1\over 2} + {1\over 2h} + \theta/2i\pi)
}\over {
\Gamma({1\over4}-\theta/2i\pi)\Gamma({1\over 4}+{1\over
2h}+\theta/2i\pi)\Gamma({1\over 2} + {1\over 2h} - \theta/2i\pi)
}} \\[0.1in]
& & \hspace{1in}\times\;\;
{{
\Gamma(1-\theta/2i\pi) \Gamma({1\over 2} -{1\over 2i\pi c} +
\theta/2i\pi) \Gamma(1-{1\over 2i\pi c} -\theta/2i\pi)
}\over{
\Gamma(1+\theta/2i\pi) \Gamma({1\over 2} -{1\over 2i\pi c} -
\theta/2i\pi) \Gamma(1-{1\over 2i\pi c} + \theta/2i\pi)
}} \;. \\
\end{eqnarray*}
The product of gamma functions alone solves (\ref{tau})
when $X_{11}=1$, and has no poles on the physical strip $0\leq {\rm
Im}\,\theta\leq\pi$; with the given $X_{11}$ we need also the $(\:)$
prefactors,
which are the reflection analogue of the CDD factors.
Possible such factors have been given by various authors\cite{toda}
investigating the (scalar) reflection factors in affine Toda
field theory. We choose those given by Fring and
K\"oberle, which are trivial at $\theta=0$ and are `minimal'
for our problem, in the sense that they introduce no
new physical poles.
Thus the only pole in $K_1(\theta)$ is that at $1/c$, which
corresponds to a boundary bound state proportional to $P^+$.

It still remains to relate $E$ to physical
boundary conditions. It is not difficult to show that
any real $N\times N$ matrix $E$ such that $E^2=I_N$ is similar
to
$$
\left( \begin{array}{c|c} I_M &  * \\ \hline
 0  & -I_{N-M}  \end{array}\right) \;.
$$
$P^-$ then leaves an $SO(N-M)$-symmetric subspace invariant,
whilst $P^+$ projects onto an $SO(M)$ subspace corresponding
to the boundary bound state at $1/c={N-2M\over 2h}i\pi$.
The boundary condition must therefore correspond to the free
condition on the $SO(N-M)$ subgroup, and a tentative
suggestion is that for the principal chiral field the
Neumann boundary condition $g'(x=0)=0$ and $g(x=0)\in SO(M)$
may be integrable.
The physics of sigma models on the half-line
remains an interesting but largely unexplored problem. In particular,
it may be interesting to investigate the Yangian
charges\cite{yangian} on the half-line.

For $M=1$ we recover Ghoshal's condition that the boundary scattering
preserve $SO(N-1)$ symmetry. We then have $c=2/i\pi$,
one of the $[\;]$ factors disappears from $\tau$ and our
expressions are those of Ghoshal, who has also, implicitly, made
the choice $K(0)=1$. Note that his $\sigma$
is rescaled by $[2][h]$ from ours because of a difference
in how we write the $S$-matrix (\ref{S}). The boundary bound
state pole is then at the edge of the physical strip, at $i\pi/2$.

\vfill\pagebreak

\section{The fused reflection matrix}

We are now in a position to fuse the $K$-matrix (\ref{K}).
The way to do this is well-known\cite{MN}: as with
$S$-matrix fusion, we make use of the fact that when
$\phi-\theta={2i\pi\over h}$, the $S$-matrix ${}_-S$ in
(\ref{reflect}) projects onto the second rank particle
and so we can replace it by $P_A+P_0$, onto which the action
of (\ref{reflect}) can now consistently be restricted.
We therefore define the reflection matrix for the second particle
to be
$$
K_2(\theta) = (P_A+P_0)\,.\, 1\otimes K(\theta+i\pi/h) \,.\,
{}_0S(2\theta) \,.\, 1\otimes K(\theta-i\pi/h)\;.
$$
The rather complicated ensuing expression for $K_2$ can eventually
be cast into quite a neat form:
$$
K_2(\theta)= \tau_2(\theta)
\left(P_A^- - [h/ci\pi-1]P_2 +[h/ci\pi-1][h/ci\pi+1] P_A^+ \right) \;,
$$
where
\begin{eqnarray*}
P^{\pm}_{A} & = &  P^\pm \otimes P^\pm \,.\, P_A \\
P_2     & = &  (P^+\otimes P^- + P^-\otimes P^+ )P_A\;+\;P_0
\end{eqnarray*}
are projectors and
$$
\tau_2(\theta) =
[1]\tau(\theta+i\pi/h)\tau(\theta-i\pi/h)\sigma(2\theta)
X_{11}(2\theta)\;.
$$
Unitarity, cross-unitarity and solution of (\ref{reflect})
are preserved by fusion and automatically follow, as does
$K_2(0)=P_A+P_0$.
This time the poles are of two types: those introduced by
the bulk $S$-matrix CDD factor (\ref{X}), and the boundary bound
states at $1/c\pm i\pi/h$. The former are a particle pole at
$\theta={i\pi\over h}$ and its crossed-channel partner.
Of the latter, certainly that at $1/c +
i\pi/h$ will be a boundary bound state, projecting onto the
antisymmetric component (only) of the $P^+\otimes P^+$ space (the
$SO(M)$-symmetric component for the $E$ discussed above): note that
at $c={2\over i\pi}$ and $M=1$ this projector vanishes, and the pole
is off the physical strip. The interpretation of the
pole at $1/c - i\pi/h$ eludes us.

\vspace{0.3in}
{\bf Acknowledgments}

I should like to thank P. Kulish, E. Corrigan, A. Macintyre,
M. Gaberdiel and P. Dorey for discussions, and the UK PPARC
for a Research Fellowship.

\pagebreak

\end{document}